\documentclass[sigconf,natbib=true,anonymous=false]{acmart}
\AtBeginDocument{%
  \providecommand\BibTeX{{%
    \normalfont B\kern-0.5em{\scshape i\kern-0.25em b}\kern-0.8em\TeX}}}



\usepackage{fancyhdr}
\renewcommand\footnotetextcopyrightpermission[1]{} 
\setcopyright{none}
\usepackage{soul}
\usepackage{url}
\usepackage{comment}
\usepackage{multirow}
\usepackage{xcolor}
\usepackage{enumitem}
\usepackage{bm}
\usepackage{verbatim}
\usepackage{amsfonts}
\usepackage{makecell}
\usepackage{dsfont}
\usepackage{balance}
\usepackage{algorithm}
\usepackage{algorithmic}

\title{A Model-based Multi-Agent Personalized Short-Video Recommender System}

\author{Peilun Zhou $\ $ Xiaoxiao Xu$^*$ $\ $ Lantao Hu $\ $ Han Li $\ $ Peng Jiang}

\begin{document}

\begin{abstract}

Recommender selects and presents top-K items to the user at each online request, and a recommendation session consists of several sequential requests. 
Formulating a recommendation session as a Markov decision process and solving it by reinforcement learning (RL) framework has attracted increasing attention from both academic and industry communities. 
In this paper, we propose a RL-based industrial short-video recommender ranking framework, which models and maximizes user watch-time in an environment of user multi-aspect preferences by a collaborative multi-agent formulization. Moreover, our proposed framework adopts a model-based learning approach to alleviate the sample selection bias which is a crucial but intractable problem in industrial recommender system.
Extensive offline evaluations and live experiments confirm the effectiveness of our proposed method over alternatives. 
Our proposed approach has been deployed in our real large-scale short-video sharing platform, successfully serving over hundreds of millions users.
\end{abstract}

\maketitle

%

\section{Introduction}
Short-video apps, such as TikTok, YoueTube Shorts, or Kuaishou, attract billions of users by creating a good user experience, which is to recommend short-form videos satisfying the user's interest.
During each user visitation to a short-video app, the recommender generates request-based recommendation, and the sequential requests forming a continuous process named session.
For short-video recommendation, taking into account the long-term reward of each recommendation action and maximizing the cumulative satisfaction of each session have become increasingly important. 
For the above aim, it has been improved to achieve more promising results to formulate a recommendation session as a Markov decision process and solving it by reinforcement learning (RL) framework.
While the multi-aspect user preferences and sample selection bias have been seldom addressed publicly, which are two crucial issues in short-video recommendation.

The user experience is usually evaluated by the user's multi-aspect feedback on the recommended videos. A user provides multi-aspect responses implying her satisfaction at each recommended video, including WatchTime (the time spent on watching the video), and several explicit interactions: Follow (follow the author of the video), Like (like this video), Comment (provide comments on the video), etc.
At our industrial short-video sharing platform, the main goal is to optimize the cumulative WatchTime over a session, and the cumulative explicit interactions are the auxiliary goals, which is similar to \cite{cai2023two}.

The existing work quite naturally takes the relation between WatchTime and explicit interactions as trade-off \cite{cai2023two, chen2021reinforcement}, however, not all explicit interactions are in competition with WatchTime. For example, if a user comments a video, she tends to spend longer time on watching this video,  etc. More collaborative relations between WatchTime and other interactions could be found. To take this issue into consideration, we propose a multi-agent recommender ranking framework, and our framework introduces a collaborative environment where multiple intelligent agent components, each maximizing distinctive user preference, work together to handle the main goal maximizing the cumulative WatchTime over a session more effectively.

Building a recommender agent offline from logged impressions with user feedback is the most practical and commonly used for the real industrial recommender, because performing online policy learning through interacting with real users usually harms user experiences. Using only logged impressions offline inevitably leads to Sample Selection Bias (SSB) \cite{ESMM2018,XuZixuan2022SSB}, which is one common but challenging issue in industrial recommender. 
Take our industrial recommender as an illustration, the ranking model makes inference on the entire space with 400 candidates, and only 6 candidates are ranked out to present and get user feedback.
Non-impression samples can also be logged, only without user feedback.
To alleviate SSB problem, we propose to extend our multi-agent framework into a model-based multi-agent framework. 
Our proposed framework alleviates SSB by introducing non-impression samples and simulating user feedback through a feedback fitting model. The feedback fitting model is optimized through iterative optimization.

In summary, the major contributions in this paper are:
\begin{itemize}[leftmargin=10pt]
\setlength{\itemsep}{-1pt}
    \item We specify a collaborative multi-agent formulation to maximize user watch-time and explicit interactions in an environment of user multi-aspect preferences.
    \item We extend the multi-agent formulation into a model-based multi-agent framework to additionally address Sample Selection Bias. 
    \item We verify the effectiveness of our proposed model-based multi-agent ranking framework through extensive experiments conducted on a public benchmark dataset and a large scale industrial dataset, and successfully apply it in a real-world large scale recommender system bringing a considerable performance improvement.
\end{itemize}

\section{The Proposed Method}


We detail the description of our method, the Model-based Multi-agent Ranking Framework (\textbf{MMRF}), which aims at maximizing user WatchTime via a multi-agent collaborative planning, and alleviating SSB by introducing simulation on non-impression samples.

\subsection{Preliminary}
Since multiple agents aim to collaboratively maximize WatchTime over a session, we formulate our problem as a multi-agent extension of Markov Decision Processes (MDPs) \cite{MDP}. Formally, we take $N$ agents, in which the $N_{th}$ agent maximizes user WatchTime, and the rest $N-1$ agents are assigned auxiliary goals for the other user preferences. \textbf{State} $S$ includes user profile, behavior history, request context and candidate item features shared among all agents. \textbf{Action} $A = [A^1, ... , A^N]$ is the action sets, in which $a^i \in A^i$ is the item score list generated by agent $i$ to determine the rank of all candidate items. \textbf{Reward} $R = [R^1, ... , R^N]$, $R^i$ is agent $i$'s reward signal, which describes an individual dimension of user's multi-aspect preference.

Our objective is to maximize the session-wise discounted total WatchTime 
$\sum_{t=0}^T{\gamma^{t} r_t^N}$, where $T$ is the time horizon, $\gamma$ is discount factor and  $r_t^N$ indicates user's WatchTime in round $t$.

\begin{figure}[t]
\centering
\includegraphics[width=0.95\linewidth]{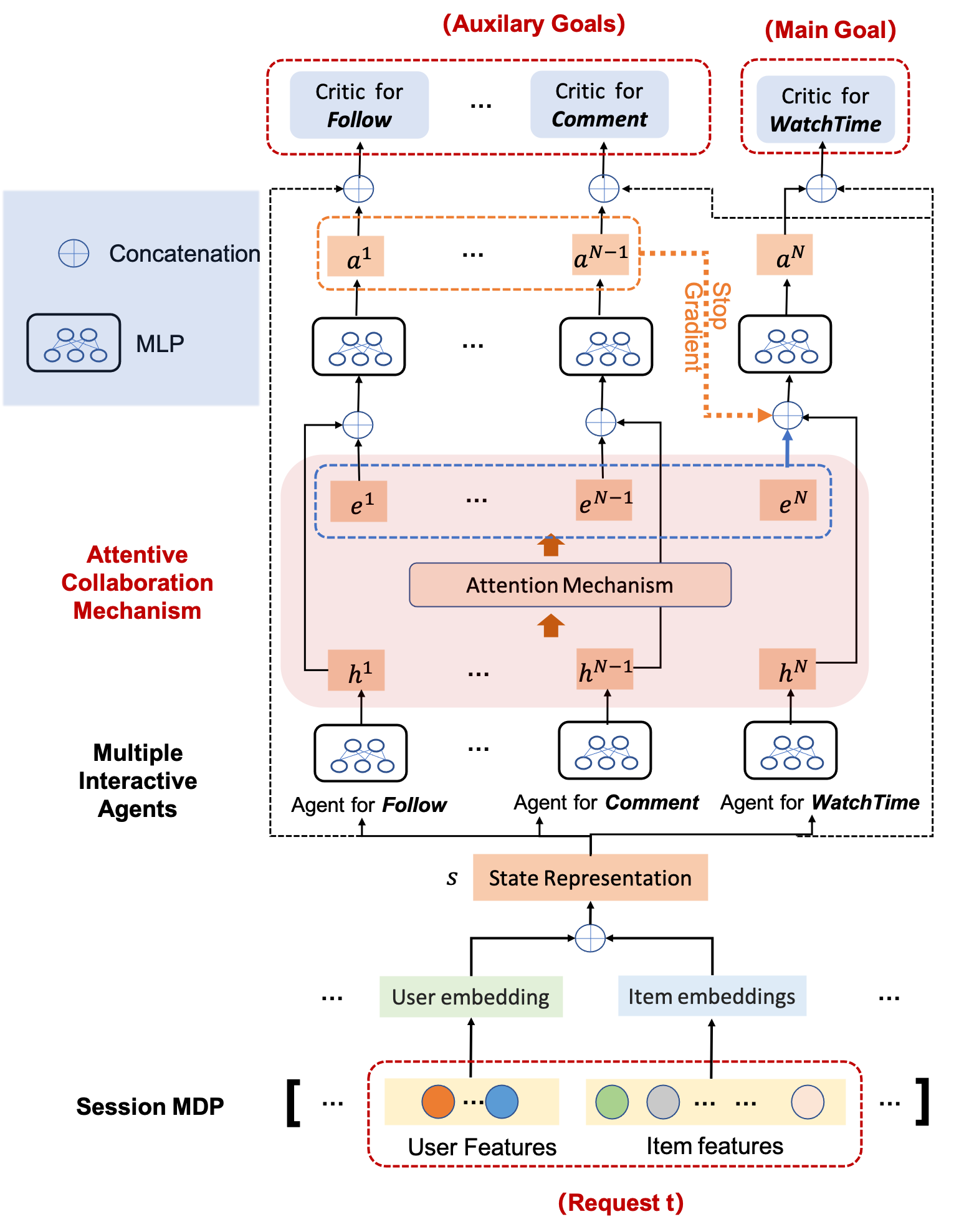}
\vspace{-2em}
\caption{An illustration of the implementation of our proposed MMRF.}
\label{main-model}
\vspace{-2em}
\end{figure}

\subsection{Multi-agent Collaboration Framework}
Since user's time spent on video watching could be motivated by other explicit interactions, there exists a collaborative relationship across multiple aspects of user preference signals. 
Based on this intuition, we propose a collaboration framework across multiple agents as illustrated in Figure \ref{main-model}. We selectively utilize knowledge from other auxiliary preference signals through attention mechanism, which is inspired by \cite{iqbal2019actor}.

\textbf{Attentive Collaboration Mechanism} 
The main idea behind our attention mechanism is to select beneficial information from other agents, so as to make a better action planning. 

Firstly, there are universal interactions among all agents by attention mechanism. For the $i_{th}$ agent, the beneficial information $e_t^i$, which is aggregated from the rest $N-1$ agents, could be formalized as:
\vspace{-0.3em}
\begin{small}
\begin{equation} \label{eq1}
\begin{aligned}
\setlength{\abovedisplayskip}{1pt}
&\bm{e_t^i} = \sum\nolimits_{j \neq i}{\alpha_j \ Relu(W_v h_t^j)}, \ \  \alpha_j \propto exp({h_t^j}^\top W_k^\top W_q h_t^i)
\setlength{\belowdisplayskip}{1pt}
\end{aligned}
\end{equation}
\end{small}
\vspace{-0.3em}
where $h_t^i$ is the state embedding of the $i_{th}$ agent. Weights $\alpha_j$ calculate the similarity between agent $i$ and $j$'s representation through multi-head attention, in which $W_q$ transform $h_t^i$ as 'query' and $W_k$ transform $h_t^j$ as 'key'. 
Notably, the weights for extracting queries, keys and values are shared across all agents, which encourages a common embedding space.

Then, with observed state $s_t$ as well as beneficial knowledge from other agents, the ranking action $a_t^k = \pi_{\theta_k}(s_t)$ of the $k_{th}$ auxiliary agent is formalized as: 
\begin{small}
\begin{equation} \label{eq10}
\begin{aligned}
\setlength{\abovedisplayskip}{1pt}
&\bm{\pi_{\theta_k}}(s_t) = f_i(h_t^k, e_t^k)
\setlength{\belowdisplayskip}{1pt}
\end{aligned}
\end{equation}
\end{small}
where $f_i(\cdot)$ adopts a multi-layer perceptron, and generates ranking scores to meet auxiliary goals.

Finally, after receiving the decision information from all the auxiliary agents, the agent for WatchTime makes its decision via action $a_t^N$ as:
\begin{small}
\begin{equation} \label{eq2}
\setlength{\abovedisplayskip}{1pt}
\bm{\pi_{\Theta}}(s_t) = f_N(h_t^N, e_t, a_t^{-N})
\setlength{\belowdisplayskip}{1pt}
\end{equation}
\end{small}
where $\Theta = \{\theta_i\}_{i=1..N}$ is a collection of $N$ actors' parameters since they need to work together to achieve the main goal. 
Not only intermediate decision knowledge $e_t = [e_t^1, ... , e_t^N]$ of multi-aspect preferences, but also direct decision results $a_t^{-N} = [a_t^1, ..., a_t^{N-1}]$ from $N-1$ auxiliary agents are integrated to support the WatchTime modeling. 

\textbf{Policy Learning}
We adopt the deterministic policy gradient (DPG) \cite{dpg} algorithm since our action space is continuous. 
In detail, for agent $i$ with actor $\pi_{\theta_i}$, considering sample $(s_t, a_t^i, r_t^i, s_{t+1})$ from replay buffer $\mathcal{D}_i$, the value function $Q_{\phi_i}$ is estimated with temporal-difference error $\delta_t^i = r_t^i + \gamma Q_{\phi_i}(s_{t+1}, \pi_{\theta_i}(s_{t+1}))- Q_{\phi_i}(s_t,a_t^i)$, and the actor $\pi_{\theta_i}$ is updated by globally maximising $Q_{\phi_i}$. 

Since we need to take advantages of auxiliary preference signals for the planning and maximising of WatchTime, the quality and accuracy of these auxiliary agents are also important. Thus, for auxiliary agents, the private auxiliary objective as well as the main objective for maximising $Q_{\phi_N}$ are both considered. As a result, the gradient updating at round $t$ could be formalized as follows:
\begin{small}
\begin{equation} \label{eq3}
\begin{aligned}
& \bm{\phi_i^{(t+1)}} \leftarrow \  \phi_i^{(t)} \ + \ \alpha_{i} \cdot \delta_t^i \cdot \nabla_{\phi_i}Q_{\phi_i}(s_t,a_t^i) \\
& \bm{\theta_i^{(t+1)}} \leftarrow \ \theta_i^{(t)} + 
\begin{cases}
\alpha_{\theta} \cdot \nabla_{\theta_i}Q_{\phi_N}(s_t,\pi_{\Theta}(s_t)), \ \ Main \ goal. \\ 
\beta_i \cdot \nabla_{\theta_i}Q_{\phi_i}(s_t,\pi_{\theta_i}(s_t)), \ \ \ Aux \ goal.
\end{cases}
\end{aligned}
\end{equation}
\end{small}

\subsection{Model-based Learning Approach with Non-impression Samples}
To address SSB problem, we introduce non-impression samples during policy learning. 
To avoid efficiency problem, we sample a subset from the complete non-impression samples randomly, and about 25\% samples are selected.
Moreover, we build a fitting model to assign user feedback for these samples.
Algorithm \ref{alg1} summarizes the details of sampling, simulating and poliy learning approach.

\begin{algorithm}
\small
\caption{The training of MMRF.}
\label{alg1}
\begin{algorithmic}[1]
\STATE \textbf{Initialization}: multi actor-critic agents parameterized by $\Theta$, $\Phi$, user feedback simulator $\mathcal{M}$ parameterized by $\eta$, replay buffer $\mathcal{D}$ and simulated buffer $\mathcal{D}^{\mathcal{M}}$; \\
\REPEAT
    \FOR{Agent $i = 1$ To $N$}
        \STATE Collecting user trajectory $\tau$ $\thicksim$ $\pi_{\theta_i}$: at round $t$, sample action $a_t^i$ according to $Eq.\ref{eq10}$ $\&$ $\ref{eq2}$, show top-K items $I_k = L_k(s_t, a_t^i)$ with user feedback $r_t^i$, next state $s_{t+1} = s_t \cup I_k$, store sample $(s_t, a_t^i, r_t^i, s_{t+1})$ in buffer $\mathcal{D}_i$;
        \FOR{$k = 1$ To $m$}
            \STATE execute random action $\widetilde{a}_t^{\ i}$ to get non-impression sample set $\widetilde{I_k} = L_k(s_t, \widetilde{a}_t^{\ i})$, obtain simulated reward $\widetilde{r}_t^{\ i}$ via user feedback simulator $\mathcal{M}$ as $Eq.\ref{eq6}$, as well as next state $\widetilde{s_{t+1}} = s_t \cup \widetilde{I_k}$, store sample $(s_t, \widetilde{a}_t^{\ i}, \widetilde{r}_t^{\ i}, \widetilde{s_{t+1}})$ in buffer $\mathcal{D}_i^{\mathcal{M}}$;
        \ENDFOR
    \ENDFOR
    \STATE \textbf{Optimising WatchTime and auxiliary goals}: Sample mini-batch from buffer $\mathcal{D}_i$ and $\mathcal{D}_i^{\mathcal{M}}$, update agent $i$, i.e., $\pi_{\theta_i}$, $Q_{\phi_i}$ via mini-batch SGD as $Eq.\ref{eq3}$;
    \STATE \textbf{Updating User Feedback Simulator} \bm{$\mathcal{M}$}: Sample mini-batch from $\mathcal{D}$, update $\eta$ via mini-batch SGD.
\UNTIL{convergence;}
\end{algorithmic}  
\end{algorithm}

Our user feedback fitting model adopts a recurrent neural network \cite{UserBehaviorModel2019}, and sequentially modeling  the transition of user state $s_t$. Since user feedback has multiple aspects, we leverage multi-head outputs, and each head is to fit a specific aspect of user feedback via minimising mean squared error:
\begin{small}
\begin{equation}
\setlength{\abovedisplayskip}{1pt}
MSE(\mathcal{M}) = ||r_t^{p_i} - r_t^i||^2
\setlength{\belowdisplayskip}{1pt}
\end{equation}
\end{small}
where $r_t^{p_i} = \mathcal{M}(s_t,a_t^i,\eta)$ is the model prediction parameterized with $\eta$. In this way, the reward of non-impression samples could be simulated. 

However, since less exposure opportunity may cause higher estimation uncertainty, those samples with high uncertainty should be more explored, so as to \textbf{reach a higher upper bound for user satisfaction}. 
Then, in order to estimate the uncertainty, inspired by \cite{rnd2018}, we extend our model to a siamese structure, with two parallel predictors $r_t^{p_i}$ and $r_t^{p_i'}$ share the same optimizing target as mentioned above. We apply dropout layers to one of the predictors and increase the discrepancy in them. In this way, the uncertainty could be estimated as the inconsistency of $P_{t}^i = (r_t^{p_i}, 1- r_t^{p_i})$ and ${P'}_{t}^i = (r_t^{p_i'}, 1-r_t^{p_i'})$. We regularize the uncertainty as KL-divergence $exp(\lambda \cdot KL(P_{t}^i||{P'}_{t}^i))$. Thus, the reward on non-impression samples could be simulated as:


\vspace{-1.0em}
\begin{small}
\begin{equation} \label{eq6}
\setlength{\abovedisplayskip}{1pt}
\widetilde{r}_t^{\ i} = mean(r_t^{p_i}, r_t^{p_i'}) \cdot exp(\lambda \cdot KL(P_{t}^i||{P'}_{t}^i))
\setlength{\belowdisplayskip}{1pt}
\end{equation}
\end{small}
The training of user feedback simulator and multiple agents are implemented in an  iterative manner as illustrated in Algorithm \ref{alg1}.

\begin{table*}[tp]
  \centering\scriptsize
  \caption{Model comparison on two datasets at seven aspects of user satisfaction. Bold font
   indicates the highest performance and * denotes the second highest performance among all baseline methods in each dimension}
  \vspace{-1.0em}
  \label{tab:rq1}
  \begin{tabular}{cc||cc||cc||cc||cc||cc||cc||cc}
    \toprule
    \multirow{10}{*}{\rotatebox{90}{KuaiRand-1k}} & \multirow{2}{*}{Methods} & \multicolumn{2}{c||}{Click$\uparrow$(1e-1)} & \multicolumn{2}{c||}{Like$\uparrow$(1e-2)} & \multicolumn{2}{c||}{Follow$\uparrow$(1e-4)} & \multicolumn{2}{c||}{Comment$\uparrow$(1e-3)} & \multicolumn{2}{c||}{Hate$\downarrow$(1e-4)} & \multicolumn{2}{c||}{LongView$\uparrow$(1e-1)} & \multicolumn{2}{c}{WatchTime$\uparrow$}\\
    & & GAUC & NCIS & GAUC & NCIS & GAUC & NCIS & GAUC & NCIS & GAUC & NCIS & GAUC & NCIS & GAUC & NCIS \\
    \specialrule{0em}{1pt}{1pt}
    \cline{2-16}
    \specialrule{0em}{1pt}{1pt}
    & BC & 0.6555 & 5.127 & 0.6586 & 1.100 & 0.6472 & 6.170 & 0.6733 & 3.084 & 0.6549 & 3.313 & 0.6130 & 2.771 & 0.6128 & 10.74
 \\
    & Wide \& Deep & 0.6603 & 5.323 & 0.6583 & 1.135 & 0.6493 & 6.175 & 0.6914 & 3.192 & 0.6886 & 3.020 & 0.6240 & 2.785 & 0.6138 & 10.71
 \\
    & DeepFM & 0.6624 & $5.368^*$ & 0.6317 & 1.086 & 0.6400 & 6.022 & 0.6807 & 3.191 & 0.6697 & 3.133 & 0.6533 & 2.933 & 0.6537 & 10.85
 \\
    & Pareto & 0.6590 & 5.227 & 0.6905 & 1.166 & $0.7056^*$ & $6.318^*$ & 0.6922 & 3.275 & $\textbf{0.7534}$ & $\textbf{2.485}$ & 0.6692 & 3.004 & 0.5947 & 9.79
 \\
    & TSCAC & $0.6633^*$ & 5.249 & $\textbf{0.7002}$ & $\textbf{1.320}$ & 0.6816 & 6.281 & $0.7253^*$ & $3.519^*$ & 0.7135 & 2.861 & 0.6616 & 2.987 & 0.6650 & 11.04
 \\
    & MASSA & 0.6238 & 4.818 & 0.6927 & 1.272 & 0.6943 & 6.297 & 0.7057 & 3.301 & 0.6126 & 3.528 & $0.6704^*$ & $3.118^*$ & 0.6679 & 11.29
 \\
    & MMRF-CO & 0.6588 & 5.328 & 0.6841 & 1.256 & 0.6909 & 6.275 & 0.6747 & 3.146 & 0.7301 & 2.811 & $\textbf{0.6734}$ & $\textbf{3.235}$ & 0.6407 & 11.03
 \\
    & MMRF-DA & 0.6601 & 5.330 & $0.6975^*$ & 1.308 & 0.6964 & 6.303 & 0.7139 & 3.383 & 0.7104 & 2.873 & 0.6695 & 2.992 & $0.6744^*$ & $11.45^*$
 \\
    & MMRF-NS & 0.6620 & 5.347 & 0.6959 & $1.312^*$ & 0.6992 & 6.311 & 0.7139 & 3.383 & 0.7302 & 2.736 & 0.6590 & 2.973 & 0.6711 & 11.35
 \\
    & MMRF & $\textbf{0.6668}$ & $\textbf{5.370}$ & 0.6837 & 1.170 & $\textbf{0.7069}$ & $\textbf{6.344}$ & $\textbf{0.7274}$ & $\textbf{3.597}$ & $0.7316^*$ & $2.795^*$ & 0.6587 & 2.972 & $\textbf{0.6833}$ & $\textbf{11.54}$
 \\
    \midrule
    \multirow{10}{*}{\rotatebox{90}{Real Production Data}} & \multirow{2}{*}{Methods} & \multicolumn{2}{c||}{Click$\uparrow$(1e-1)} & \multicolumn{2}{c||}{Like$\uparrow$(1e-2)} & \multicolumn{2}{c||}{Follow$\uparrow$(1e-3)} & \multicolumn{2}{c||}{Comment$\uparrow$(1e-2)} & \multicolumn{2}{c||}{Hate$\downarrow$(1e-4)} & \multicolumn{2}{c||}{longView$\uparrow$(1e-1)} & \multicolumn{2}{c}{WatchTime$\uparrow$}\\
    & & GAUC & NCIS & GAUC & NCIS & GAUC & NCIS & GAUC & NCIS & GAUC & NCIS & GAUC & NCIS & GAUC & NCIS\\
    \specialrule{0em}{1pt}{1pt}
    \cline{2-16}
    \specialrule{0em}{1pt}{1pt}
    & BC & 0.6587 & 5.660 & 0.6570 & 2.947 & 0.5981 & 1.933 & 0.6206 & 1.095 & 0.6570 & 3.500 & 0.7279 & 2.875 & 0.6378 & 25.18
 \\
    & Wide \& Deep & 0.6568 & 5.743 & 0.6544 & 3.003 & 0.5961 & 1.951 & 0.6190 & 1.066 & 0.6551 & 3.439 & 0.7296 & 2.903 & 0.6413 & 25.93
 \\
    & DeepFM & 0.6862 & 5.923 & 0.6484 & 2.987 & 0.6045 & 1.965 & 0.6215 & 1.182 & 0.6868 & 3.269 & $0.7598^*$ & $3.063^*$ & 0.6501 & 26.01
 \\
    & Pareto & 0.6910 & 6.035 & 0.6508 & 2.913 & 0.6241 & 1.995 & 0.6548 & 1.156 & 0.6932 & 2.951 & 0.7422 & 3.033 & 0.6476 & 25.73
 \\
    & TSCAC & 0.7008 & 6.111 & 0.6459 & 2.829 & 0.6198 & 1.931 & 0.6557 & 1.189 & 0.7037 & 2.914 & 0.7374 & 2.958 & 0.6510 & 26.18
 \\
    & MASSA & 0.6885 & 5.845 & $\textbf{0.6685}$ & $\textbf{3.197}$ & 0.6309 & 2.1230 & $0.6573^*$ & $1.227^*$ & 0.6918 & 3.110 & 0.7399 & 2.954 & 0.6466 & 24.26
 \\
    & MMRF-CO & $0.7139^*$ & $6.240^*$ & 0.6240 & 2.505 & $0.6367^*$ & $2.134^*$ & 0.6417 & 1.196 & 0.7110 & 2.939 & 0.7580 & 3.038 & 0.6490 & 24.78
 \\
    & MMRF-DA & 0.6957 & 6.079 & 0.6626 & 3.025 & 0.5971 & 1.913 & 0.6308 & 1.111 & 0.6968 & 2.934 & $\textbf{0.7721}$ & $\textbf{3.255}$ & $\textbf{0.6595}$ & $26.85^*$
 \\
    & MMRF-NS & 0.7018 & 6.120 & $0.6634^*$ & $3.030^*$ & 0.6078 & 1.976 & 0.6232 & 1.191 & $0.7094^*$ & $2.903^*$ & 0.7538 & 3.001 & 0.6579 & 26.61
 \\
    & MMRF & $\textbf{0.7192}$ & $\textbf{6.270}$ & 0.6421 & 2.889 & $\textbf{0.6776}$ & $\textbf{2.389}$ & $\textbf{0.6719}$ & $\textbf{1.287}$ & $\textbf{0.7192}$ & $\textbf{2.875}$ & 0.7564 & 2.989 & $0.6585^*$ & $\textbf{26.89}$
 \\
    \bottomrule
  \end{tabular}
\vspace{-1em}
\end{table*}

\section{Experiments}
\subsection{Offline Experiments}
\subsubsection{Datasets} 
Offline experiments are conducted in an online simulator of one public dataset and one real production dataset.
\textbf{Public Dataset: }
The public dataset KuaiRand1K \cite{gao2022kuairand} is collected from a famous video-sharing mobile app and suitable for the offline evaluation of RL methods. We adopt the data pre-processing approach in \cite{cai2023two}, which chronologically concatenates logs of the same user to form a trajectory. 
\textbf{Real Production Dataset: }
This dataset are traffic logs collected over a week from our large scale video-sharing platform including 245 billion interactions between 1.8 billion users and 1.0 billion videos. Chronologically concatenated logs of the same session form a trajectory.

\subsubsection{Evaluation} 
We user the \textit{Normalised Capped Importance Sampling} (\textbf{NCIS}) approach to evaluate different policies, which is a standard offline evaluation approach for RL-based methods in recommender systems \cite{zou2019reinforcement}. We also evaluate our method in terms of \textbf{GAUC}, which is a commonly used metric in recommendation ranking models \cite{zhou2018deep}.

\subsubsection{Baselines} 
We compare our approach with: 1) \textbf{Industrial commonly applied approaches}, including supervised behavior-cloning (\textbf{BC}) policy $\pi_{\beta}$, {\textbf{Wide\&Deep} \cite{WideandDeep2016} and \textbf{DeepFM} \cite{dfm2017} which are supervised by impression signals, samples are weighted via the weighted sum of user feedback. 
2) \textbf{Multi-objective/multi-agents RL approaches}, including \textbf{Pareto} \cite{ChenXu2021} as a multi-objective RL algorithm that finds the Pareto optimal solution for recommender systems, \textbf{TSCAC} \cite{cai2023two} as a two-stage multi-critic algorithm that maximising main response and satisfy the constraints of other auxiliary responses, and \textbf{MASSA} \cite{HEXu2020RL} as a multi-agent algorithm, in which a signal network is adopted to handle messages sent by multiple critics and extract useful information to each critic.
3) \textbf{Several alternatives of our framework MMRF}, including \textbf{MMRF-DA} without learning from non-impression samples, \textbf{MMRF-NS} in which the adoption of user feedback simulator is replaced by a heuristic method, i.e., treating the non-impression samples with a negative constant reward, and \textbf{MMRF-CO} with no attentive collaboration mechanism but a simple concatenation.

\subsubsection{Experimental Results} 
Table~\ref{tab:rq1} presents the comparison between our proposed MMRF and other alternatives. 

We can see that our MMRF algorithm significantly outperforms the others: 1) For the main goal (WatchTime), MMRF achieves the highest performance with 7.3\% GAUC improvement and 7.1\% NCIS improvement. 2) For auxiliary goals, MMRF also achieves top2 performance for 4 out of 6 dimensions. The Pareto algorithm indeed learns a pareto optimal solution with no obvious shortcomings in all dimensions, but it does not satisfy our main goal and its performance on WatchTime is very average (-0.7\% on GAUC). Although TSCAC and MASSA adopt multi-critic/multi-agent structure, due to the lack of communication across agents, it seems to easily fall into sub-optimal.

The results of ablation study are also presented in Table~\ref{tab:rq1}.
We first investigate the effect of collaboration. Note that MMRF outperforms MMRF-CO for 5 out of 7 satisfaction aspects, which confirms that the collaboration mechanism indeed achieves comprehensive improvement.
We also explore the effect brought by utilizing non-impression samples. Comparison between MMRF-DA and MMRF shows that as a data argument method, the imitation of user feedback simulator could benefit the learning on sparse signals, with more than 12.7\% NCIS improvement at follow and 2.3\% NCIS improvement at hate. However, MMRF-NS which treats non-impression samples with a negative constant reward shows no obvious advantages compared with MMRF-DA. This phenomenon indicates that simply introducing non-impression samples and treating them equally could not effectively address SSB issue.

\subsection{Online A/B Testing}
To demonstrate the effectiveness of our proposed framework, we test its performance as well as other alternatives via A/B testing in our large scale industrial video-sharing platform. In our platform, when a user arrives, these algorithms are expected to rank the candidates, and the top videos will be recommended to the user.

\subsubsection{Setup} 
Here we explain our adopted online evaluation metrics and experimental details.
\textbf{Online Evaluation Metrics: }
For the main goal, we look at the averaged total amount of time user spend on the videos, referred to as WatchTime. 
The total number of the viewed videos among a session (Depth for short) and the the total number of explicit interactions (Follow and Comment) can also speak of user satisfaction.  
\textbf{Experimental Details: }
We complement our evaluation with a supervised learning-to-rank (LTR) \cite{liu2009learning} baseline, and take \textbf{TSCAC} as the alternative.
To test the above threee algorithms, we randomly split users on the platform into three backets. The first bucket runs the baseline LTR model, and the remaining buckets run our proposed method and TSCAC. Models are trained for a month and then fixed to test performance within the following one day.

\begin{table}[H]
  \centering\small
  \caption{Performance comparison of our MMRF and the STOA TSCAC with the LTR baseline in online A/B testing.}
  \vspace{-1em}
  \label{ab-test}
  \renewcommand\arraystretch{1.0}
    \begin{tabular}{l|c|c|c|c}
    \hline
    Online A/B Test & WatchTime & Depth & Follow & Comment\\
    \hline
    LTR & - & - & - & -\\
    TSCAC & +0.32\% & +0.31\% & +0.69\% & +0.42\%\\
    MMRF & +0.55\% & +0.54\% & +1.45\% & +1.28\%\\
    \hline
    \end{tabular}
  \vspace{-1em}
\end{table}

\subsubsection{Results Analysis} 
Table \ref{ab-test} shows the performance improvement of algorithm comparison with the LTR baseline regarding metrics WatchTime, Depth, Follow and Comment. 
As we can see, our proposed method contributes higher performance improvement than TSCAC which has already been highly optimized for recommender system. 
Note that 0.5\% increase in WatchTime is critial for our full-fledged video-sharing platform.
Figure \ref{online-period} plots the online performance improvement of our proposed algorithm over the LTR baseline over a sustained 10 days. Note that MMRF is deployed at the end of Day3.

\begin{figure}[t]
\centering
    \begin{minipage}[t]{0.5\linewidth}
      \centering
      \centerline{\includegraphics[width=0.99\linewidth]{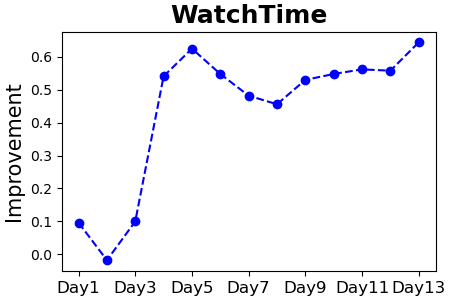}}
      \vspace{-0.5em}
      \centerline{\footnotesize{(a)}}
      \centering
    \end{minipage}%
    \begin{minipage}[t]{0.5\linewidth}
      \centering
      \centerline{\includegraphics[width=0.99\linewidth]{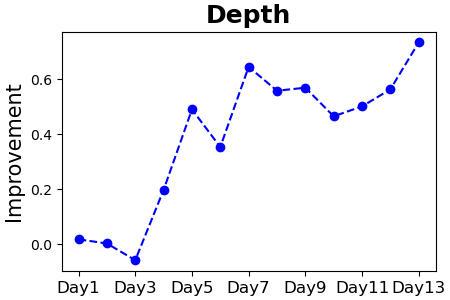}}
      \vspace{-0.5em}
      \centerline{\footnotesize{(b)}}
      \centering
    \end{minipage}%
    \hspace{0.1em}
    \begin{minipage}[t]{0.5\linewidth}
      \centering
      \centerline{\includegraphics[width=0.99\linewidth]{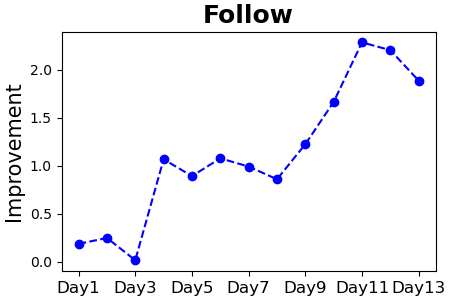}}
      \vspace{-0.5em}
      \centerline{\footnotesize{(c)}}
      \centering
    \end{minipage}%
    \begin{minipage}[t]{0.5\linewidth}
      \centering
      \centerline{\includegraphics[width=0.99\linewidth]{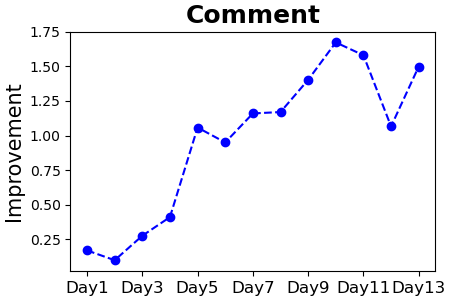}}
      \vspace{-0.5em}
      \centerline{\footnotesize{(d)}}
      \centering
    \end{minipage}%
  \vspace{-1.0em}
  \caption{Online performance improvement of our proposed MMRF over the LTR baseline of each day.}
  \label{online-period}
\vspace{-1.5em}
\end{figure}

\section{Conclusions}
We propose MMRF to optimize the cumulative WatchTime and multi-aspect explicit interactions over a session in short video recommendation.
MMRF maximizes the cumulative WatchTime over a session efficiently and effectively by introducing a collaborative multi-agent environment.
Moreover, MMRF addresses Sample Selection Bias through introducing non-impression samples and building a feedback fitting model to simulate user feedback for non-impression samples.
Extensive offline experiments and online A/B Testing are conducted and the results demonstrate the effectiveness of our proposed method.
In fact, MMRF has been deployed in our real-world large-scale video-sharing platform and successfully serving over hundreds of millions of users. 

\bibliographystyle{ACM-Reference-Format}
\balance
\bibliography{main}

\end{document}